\documentclass[aps,twocolumn,showpacs,preprintnumbers,floats]{revtex4}
\usepackage{amsfonts}
\usepackage{amsmath}
\usepackage{graphicx}
\usepackage{dcolumn}
\usepackage{bm}
\usepackage{color}

\begin{document}

\preprint{}
\title{Ferromagnetism and increased ionicity in epitaxially grown TbMnO$_3$ films}
\author{D. Rubi$^{1}$, C. de Graaf$^{2}$, C. J. M. Daumont$^{1}$, D. Mannix$^{3}$, R. Broer$^{1}$, B. Noheda$^{1}$}
\email{b.noheda@rug.nl}
\affiliation{$^{1}$Zernike Institute for
Advanced Materials, University of Groningen, Groningen 9747AG, The
Netherlands}
\affiliation{$^{2}$Instituci\'o Catalana de Recerca i Estudis Avan\c{c}ats (ICREA), 08010 Barcelona, Spain}
\affiliation{$^{3}$Institut N$\acute{e}$el, CNRS-UJF, 25 Avenue des Martyrs, 38042 Grenoble Cedex 9, France}
\date{October 20, 2008}

\begin{abstract}
Thin films of TbMnO$_3$ have been grown on SrTiO$_3$ substrates. The films grow under compressive strain and are only partially clamped to the substrate. This produces remarkable changes in the magnetic properties and, unlike the bulk material, the films display ferromagnetic interactions below the ordering temperature of $\sim$40K. X-ray photoemission measurements in the films show that the Mn-3s splitting is 0.3 eV larger than that of the bulk. \textit{Ab initio} embedded cluster calculations yield Mn-3s splittings that are in agreement with the experiment and reveal that the larger observed values are due to a larger ionicity of the films.
\end{abstract}

\maketitle

Orthorhombic TbMnO$_3$ is a multiferroic\cite{Hil00,Eer06}, that is, it displays both antiferromagnetic and ferroelectric order at low temperatures\cite{Fie05,Kim03}. Moreover, these two ferroic orderings are so strongly coupled that the electrical polarization can be flipped by a magnetic field. This large magnetoelectric coupling has recently made of TbMnO$_3$ a very popular material. However, films of orthorhombic TbMnO$_3$ have seldom been reported.

From a fundamental viewpoint, a main advantage of using thin films of TbMnO$_3$ is the possibility of modifying the structure using the strain imposed by the substrate. This approach should allow for extra degrees of freedom compared to the well-established method of changing the Mn-O-Mn bond angle by rare earth substitution and should shed light into the structural details that determine the magnetic and ferroelectric behavior. On the application side, thin films of TbMnO$_3$ are also of clear interest where integration and miniaturization is required. Moreover, the possibility of tuning the exchange interactions and inducing ferromagnetic behavior via strain could lead to novel single-phase ferrimagnetic ferroelectrics, which are highly interesting for applications and very scarce\cite{Hil00}.

Bulk TbMnO$_3$ (TMO) displays a complex magnetic behavior. A first antiferromagnetic transition takes place at T$_N$$\sim$40K -where Mn spins order in a sinusoidal incommensurate structure-. As the temperature is further decreased, the propagation vector of the sinusoidal structure is reduced until it locks at T$_{lock}$ $\sim$28K, where the magnetic structure changes to a spiral antiferromagnetic ordering. A spontaneous electrical polarization (P$_s$) along the c-axis and a strong magnetoelectric (ME) effect are observed below T$_{lock}$\cite{Kim05}. Using symmetry considerations, Kenzelman et al.\cite{Ken06} and Mostovoy \cite{Mos06} have shown that a P$_s$ must exist in any spiral magnet. Due to this direct relationship between the magnetic structure and P$_s$, the ME coupling in these materials is very strong. An electronic origin was first reported to explain the ferroelectric polarization observed in TbMnO$_3$\cite{Kat05}. However, Sergienko et al \cite{Ser06} have proposed that the Dzyaloshinskii-Moriya interaction is the microscopic mechanism responsible for such effect. Vanderbilt et al.\cite{Mal08} and Xiang et al.\cite{Xia08} have recently confirmed that ionic displacements are indeed at the origin of the ferroelectricity in this material.

The few existing reports on thin films of orthorhombic TbMnO$_3$ are on relatively thick (relaxed) films, for which epitaxial strain does not play a clear role\cite{Cui05}. Here we have used epitaxial strain to modify the structure of TbMnO$_3$. This has been done by growing the films with small enough thicknesses on SrTiO$_3$ substrates. We show that the strained films indeed differ substantially from their bulk counter parts. The films display ferromagnetic-like interactions below the bulk Neel temperature. Moreover, an increased XPS Mn3s splitting is observed, which is found to be directly related to an increase in the ionicity of the films with respect to the bulk structure.

(001)-oriented TbMnO$_3$ (TMO) thin films were deposited on atomically flat TiO$_2$-terminated (001)-SrTiO$_3$ (STO) cubic substrates by Pulsed Laser Deposition. The TMO deposition was performed at 750$^o$C at oxygen pressures ranging from 0.25 to 0.9 mbar. Structural characterization, including high resolution synchrotron measurements, shows that the thin films have a distorted perovskite structure free from secondary phases\cite{Dau08}. The thin films are clamped to the substrate along one of the in-plane [100]-directions while they maintain an orthorhombic structure, as shown in Figure 1(a). This results in that, despite the partial clampling, the films are strained along both pseudo-cubic directions. The four equivalent orientational domains, all with the c-axis out of plane, are present so the films keep the four-fold macroscopic symmetry of the substrate (see Figure 1(b)).

\begin{figure}
	\centering
\includegraphics[scale=0.4]{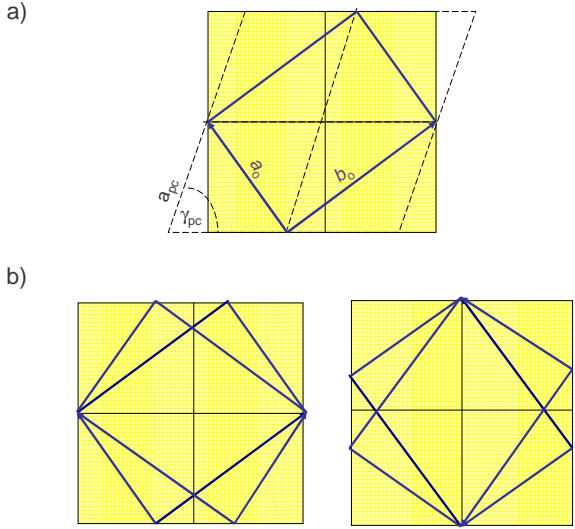}
\caption{(Color online)(a) Sketch of the TbMnO$_3$ orthorhombic unit cell, the pseudo-cubic TbMnO$_3$ lattice (dashed lines) and the cubic substrate (shaded). (b) Sketch of the four types of orthorhombic domains present in the films}
\end{figure}

Figure 2(a) shows the orthorhombic cell parameters for films of different thicknesses grown under an oxygen pressure of 0.9mbar, as obtained by X-ray diffraction area maps of the reciprocal space around the (113) and (103) substrate reflections. In Figure 2(b), the corresponding pseudo-cubic lattice parameters are plotted. In this figure it is clearly seen that the strained films share the pseudo-cubic lattice parameters of the substrate and are, therefore, compressed in the plane directions from about 3.94 {\AA} of the bulk to 3.90{\AA}. The orthorhombic distortion, represented by the deviation from 90$^o$ of the pseudo-cubic angle $\gamma_{pc}$, is smaller than the bulk one. This gives rise to a highly compressed ${b_o}$ and a slightly expanded ${a_o}$ axis. The out-of plane ${c_o}$ lattice parameter is enlarged in the films, consistent with an overall compressive in-plane strain.

\begin{figure}
	\centering
\includegraphics[scale=0.4]{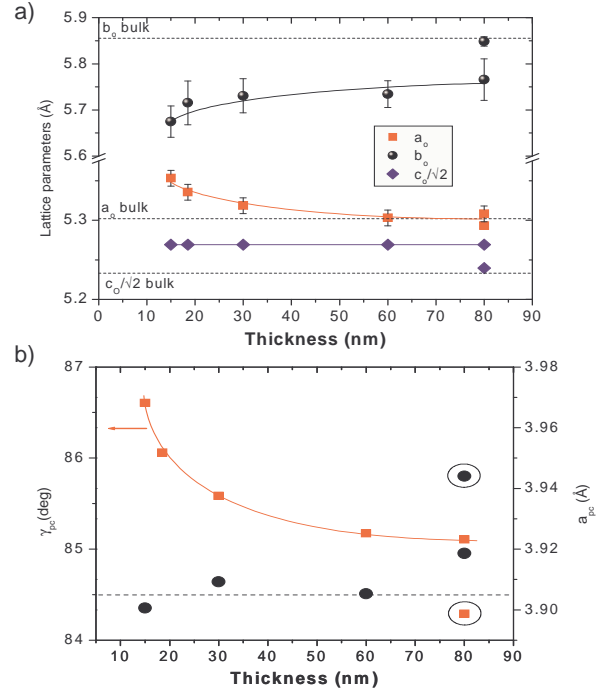}
\caption{(Color online)Orthorhombic (a) and pseudo-cubic (b) lattice parameters of a series of TbMnO$_3$ films with different thickness grown at PO$_2$=0.9 mbar. Encircled are the bulk-like values corresponding to the relaxed part of the 80nm film.}
\end{figure}

Upon increasing thickness ($d$) the structure changes in an unconventional way: The out-of-plane cell parameter, $c_o$ remains constant with thickness, indicating that the strain remains constant. The orthorhombic distortion increases while $a_o$ decreases approaching the bulk value. At the same time $b_o$ increases but, unlike $a_o$, it approaches a value substantially smaller than that of the bulk. This gradual change is maintained at least until $d$= 60nm and at $d$= 80nm, part of the film undergoes a sudden relaxation to the bulk structure\cite{Alo00} (dashed lines in Fig.2(a)). The crystal structure of the films grown at 0.25 mbar shows the same trend but slightly larger $c_o$, likely due to the presence of oxygen vacancies\cite{Dau08}. The presence of oxygen vacancies in the films grown at 0.9 mbar is unlikely since the bulk unit cell is reproduced for the relaxed samples.

 \begin{figure}[tbp]
 \centering
\includegraphics[scale=0.35]{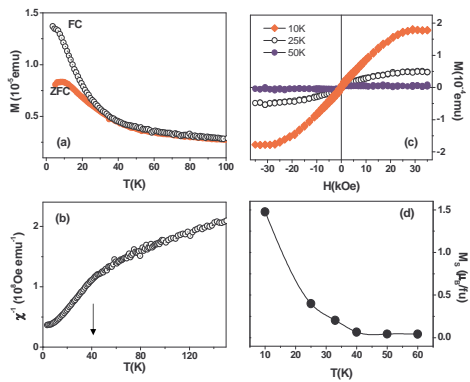}
\caption{(Color online)(a) Magnetization (M) as a function of temperature (T) for a 40nm TMO film grown at 0.9mbar. Data were recorded upon warming, with a 500 Oe field applied parallel to the film's surface, altough qualitatively similar results were obtained when the field was applied in the perpendicular direction. The diamagnetic contribution arising from the STO substrate was measured in a separate control experiment and subtracted from the raw magnetization. (b)Inverse susceptibility ($\chi^{-1}$) as a function of T. (c) M versus magnetic field (H) at different temperatures. Both the diamagnetism of the substrate and the paramagnetism of Tb-ions have been deducted from the raw data. (d) Saturation magnetization (M$_S$) as a function of T.}
\end{figure}

Figure 3(a) shows the magnetization as a function of temperature for a 40nm TMO film grown at 0.9mbar, measured under zero-field-cooling (ZFC) and field-cooling (FC) conditions. The magnetization presents an up-turn at T*$\sim$40K. This feature can be better appreciated in the evolution of the inverse susceptibility ($\chi^{-1}$) with temperature (Figure 3(b)), where a change of slope can be seen at T*. The modeling of the high temperature tail of Figure 3(b) by means of a Curie-Weiss law gives a negative extrapolated temperature ($\theta_{CW}\sim$ -150K), indicating that the dominant magnetic interaction is antiferromagnetic. This suggests that the transition observed at T* is ferrimagnetic-like. The splitting at low temperatures between FC and ZFC measurements - typical of a glassy behavior - indicates that ferromagnetic interactions are present in the films. Figure 3(c) shows magnetization loops measured at 10, 25 and 50K and confirms the presence of ferromagnetism. Figure 3(d) depicts the evolution of the saturation magnetization (M$_S$) as a function of temperature, showing that ferromagnetism develops below T*$\sim$40K. Similar effects have recently been observed in orthorhombic YbMnO$_3$ and YMnO$_3$ thin films \cite{Mar07,Rub08}, which points towards a general mechanism in manganite thin films.

We notice that the observed magnetic ordering temperature ($\sim$40K) is very close to the transition temperature to the sinusoidal antiferromagnetic structure in the bulk compound; however, the weak ferromagnetism appearing in our films has not counterpart in bulk. Moreover, our magnetic measurements did not reveal any feature related to the stabilization of the spiral antiferromagnetic ordering -and the concomitant onset of ferroelectricity. Special attention should be paid to the possible existence of Mn$_3$O$_4$ impurities (T$_C$=42K)\cite{Man04})(even though we were unable to observe them by x-ray diffraction), which could account for the presence of ferromagnetism at low temperatures. However, the measured ferromagnetic saturation (M$_S$$\sim$1.5 $\mu_B$/f.u at 10K) is higher that the saturation magnetization expected for Mn$_3$O$_4$ M$_S$$\sim$0.5 $\mu_B$/Mn\cite{Man04}, strongly indicating that the latter cannot account for the observed ferromagnetism. The magnetic characterization corresponding to films with different thickness and grown at very different oxygen pressures (0.25 and 0.9 mbar) displayed analogous results.

Figure 4(a) shows the X-ray photoemission spectroscopy (XPS) spectra at the Mn3s edge corresponding to films grown at oxygen pressures of 0.25 and 0.9 mbar. The Mn-3s level splitting originates in the intra-atomic exchange coupling between 3s and 3d electrons and the magnitude of the splitting is reported to increase linearly as the local ionic Mn valence decreases\cite{Gal02,Dor06}. The maximum experimental values reported for Mn ions with a 3+ nominal valence is 5.3eV \cite{Gal02}. Figure 4(a) shows that in our films the splitting is 5.7(1)eV.  This value is independent of the thickness and the oxygen pressure during growth. It can be suggested that the enlarged splitting is due to a mixed +2/+3 valence \cite{Gal02} that could be originated by oxygen vacancies (a likely defect in these compounds). However, the absence of shake-up peaks in the Mn2p spectra (see Fig.4(b)) goes against the presence of Mn$^{2+}$ in the films. Moreover, the splitting remains the same (within the accuracy of the setup) no matter the oxygen growth pressure. The presence of oxygen vacancies has thus no noticeable influence in the observed splitting.

\begin{figure}[tbp]
 \centering
\includegraphics[scale=0.4]{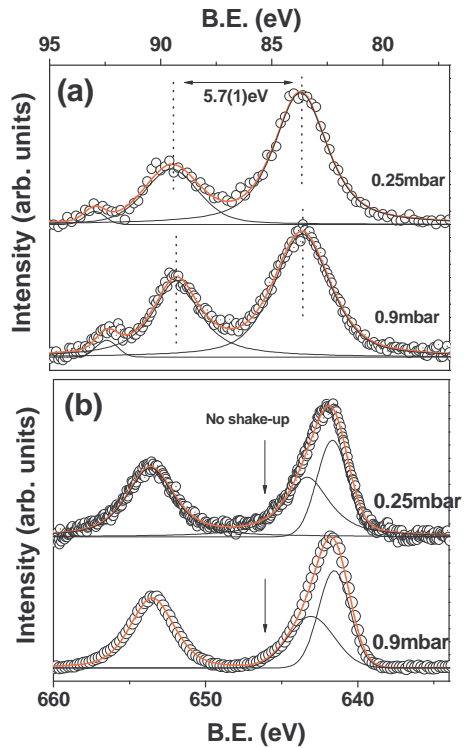}
\caption{(Color online)XPS spectra around the Mn3s (a) and Mn2p (b) edges for two films grown at two very different oxygen pressures. The observed spectra are independent of oxygen pressure during growth and film thickness. The small peak appearing at 93eV is likely to be related to an unidentified surface contamination.}
\end{figure}

Therefore, the increased Mn3s splitting in the films should be explained in terms of a more complex scenario. In order to shed light into this problem we have performed configuration interaction (CI) calculations within the embedded cluster approach. The electronic structure of an MnO$_6$ cluster is calculated with accurate quantum chemical schemes that ensure a precise and unbiased treatment of the strong electron correlation effects present in this type of materials. This MnO$_6$ cluster is embedded in a set of point charges that reproduce the Madelung potential in the cluster region due to the rest of the crystal. To avoid an artificial delocalization of the cluster electrons to the point charges, the centers nearest to the cluster are represented with model potentials \cite{Bar88} that account for the Coulomb and exchange interactions between the electrons in the cluster and the surroundings. This local approach has been successfully applied in the past to interpret XPS spectra of various ionic TM oxides \cite{Gra97,Hoz01}.

The final states responsible for the two peaks in the Mn3s XPS spectrum are characterized by a 3s$^1$3p$^6$3d$^5$ electronic configuration. Only taking into account this configuration, very large exchange splittings are obtained in the calculations. However, it was shown by Bagus and collaborators \cite{Bag04} in the analysis of the exchange splitting in MnO that important contributions to the wave function arise from 3s$^2$3p$^4$3d$^6$ electronic configurations. Semi-quantitative agreement with experiment can be obtained by also including the 3s$^2$3p$^5$3d$^4$4f$^1$ configurations. Applying this strategy to embedded MnO$_6$ clusters representing MnO, LaMnO$_3$ and CaMnO$_3$ give exchange splittings of 6.38, 5.72 and 3.95 eV, respectively. These calculated values are in good agreement with the experimental numbers: 6.2 eV for MnO; 5.3 eV for LaMnO$_3$ and 4.0 eV for CaMnO$_3$.

The embedded cluster for bulk TbMnO$_3$ was constructed using the experimental structure \cite{Alo00}. For the thin film cluster, we applied the lattice parameters of the thinnest film reported in Fig 2(a). The CI calculations give an exchange splitting of 5.18 eV for bulk TbMnO$_3$ and 5.46 eV for the thin film. The slight underestimation in comparison to the experimental value of 5.7 eV reported in Fig. 4(a) is to be expected because in the calculations the 4f type expansion functions were not optimized. The increase of +0.3 eV in comparison to bulk is precisely what is observed in the experiment.

The steady increase of the exchange splitting from CaMnO$_3$ to LaMnO$_3$ to MnO suggests that the exchange splitting is determined by the formal ionic Mn charge as reported by Galakhov\cite{Gal02}. However, this relationship cannot be used to explain the different exchange splitting in bulk and thin film TbMnO$_3$, since in both cases the formal Mn charge is the same. Actually the calculated Mn charge is slightly smaller in the bulk than in the thin film contradicting the suggested relationship between Mn charge and exchange splitting. Serious doubts have been raised on the usefulness of the concept of the formal charge and/or oxidation state to interpret the electronic structure of transition metal compounds \cite{Rae08}. Instead, we analyze the relation between the exchange splitting and the screening of the core hole by the oxygen ligands. For this purpose, the $N$-electron wave function is expressed in localized orbitals \cite{Sad07} and configurations are grouped by non charge transfer (Mn-3d$^5$), charge transfer (Mn-3d$^6$L$^{-1}$) and configurations with  two or more electrons transferred from oxygen to Mn. This analysis shows that the screening of the core hole by the oxygens is more effective in the bulk than in the film; the charge transfer (CT) configurations have a larger weight in the wave function of the bulk cluster (44\%) than in the film (38\%). Hence, instead of the formal Mn charge, the exchange splitting is determined by the degree of oxygen screening. It is well known that MnO is highly ionic with almost no charge transfer in the wave function, while CaMnO$_3$ has a more covalent character with a much larger degree of  screening, in line with the observed exchange splittings.

The reduced degree of core hole screening by the oxygens in the films is related to the shorter Mn-O distances in the ab-plane. These shorter distances make that the oxygen-to-metal charge transfer configuration lies higher in energy and contributes less to the wave function. In a one-electron reasoning this can be explained by the fact that shorter Mn-O distances lead to enhanced anti-bonding interactions, which increase the Mn-3d orbital energies. This causes a higher charge transfer energy, and hence, less effective ligand screening.

It is tempting to say that the decreased charge transfer in the films reduces the efficiency of the superexchange mechanism and enhances ferromagnetism with respect to the bulk case, in agreement with our experimental observations. However, given the subtle competition between ferromagnetic and antiferromagnetic exchange constants in this material\cite{Xia08} and the likely influence of the Tb ions, a more complex analysis of the magnetic structure of the films is needed to explain the observed ferromagnetism. For that, the full structure determination of the films, including the oxygen atomic positions, is compulsory.

Another possible origin of the observed ferromagnetism is the coupling between magnetization and strain. In this respect, there is an analogy between the induction of ferromagnetism in epitaxial antiferromagnets and the well-known induced ferroelectricity in incipient ferroelectrics\cite{Pert00,Bell07}. Indeed, antiferromagnets are piezomagnetics and the epitaxial strain should induce a magnetic moment in the films. Moreover, the linear part of the magnetostriction in antiferromagnets is usually several orders of magnitude larger\cite{Newn}. The strong coupling of the magnetic structure of TbMnO$_3$ to the lattice has been recently demonstrated\cite{Mele07}. In our films, the epitaxial stress is estimated to be 2x10$^8$ N/m$^2$ (using a value of the Young modulus of 20 GPa\cite{Lali08}) and, thus, the magnetization values observed are compatible with an effective piezomagnetic coefficient of 10$^{-10}$ m/A. This value is substantially smaller than the reported magnetostriction bulk values\cite{Mele07} and one order of magnitude larger than typical intrinsic piezomagnetic coefficients in antiferromagnets\cite{Newn}. In addition, the presence of in-plane domains in the films also makes it difficult to extract any final conclusions from these values.

In summary, TbMnO$_3$ films grown epitaxially on SrTiO$_3$ substrates display a strained orthorhombic perovskite structure less distorted than that of the bulk. This structural modification gives rise to an increased ionicity in the films and to very different magnetic properties: The films show ferromagnetism below $\sim$40K and do not display signs of a second low temperature magnetic transition, as the one associated to the onset of ferroelectricity in bulk TbMnO$_3$.

The authors are grateful to G. Catalan, J. Fontcuberta, X. Marti, M. Mostovoy and T.T. M. Palstra for very useful discussions. We would like to thank T.F. Landaluce and P. Rudolf for the assistance with and access to the XPS measurements. This work was supported by the E.U. STREP MaCoMuFi (Contract FP6-NMP3-CT-2006-033221) and the Spanish ministry of science (Grant CTQU2005-08459-C02-02/BQU).

\end{document}